\documentclass[runningheads]{llncs}
\usepackage{todonotes}
\usepackage{graphicx}
\usepackage[T1]{fontenc} 
\usepackage[utf8]{inputenc}
\usepackage{multirow}
\begin{document}
\title{A Pilot Study of a Human-Readable \\Robotic Process Automation Language}
\titlerunning{A Pilot Study of a Human-Readable RPA Language Proposal}

\author{Piotr~Gago\inst{1}\orcidID{1111-2222-3333-4444} \and
Daniel~Jablonski\inst{1}\orcidID{0000-1111-2222-3333} \and
Anna~Voitenkova\inst{1}\orcidID{1111-2222-3333-4444} \and
Ihor~Debelyi\inst{1}\orcidID{1111-2222-3333-4444} \and
Kinga~Skorupska\inst{1}\orcidID{0000-0002-9005-0348} \and
Maciej~Grzeszczuk\inst{1}\orcidID{0000-0002-9840-3398} \and
Wieslaw~Kopec\inst{1}\orcidID{0000-0001-9132-4171}} 

\authorrunning{Gago et al.}
\institute{Polish-Japanese Academy of Information Technology}

\maketitle              % typeset the header of the contribution
\begin{abstract}
In this paper, we explore the usability of a custom eXtensible Robotic Language (XRL) we proposed. To evaluate the user experience and the interaction with the potential XRL-based software robot, we conducted an exploratory study comparing the notation of three business processes using our XRL language and two languages used by the leading RPA solutions. The results of our exploratory study show that the currently used XML-based formats perform worse in terms of conciseness and readability. Our new XRL language is promising in terms of increasing the readability of the language, thus reducing the time needed to automate business processes.

\end{abstract}

\begin{figure*}[h!]
  \includegraphics[width=\textwidth]{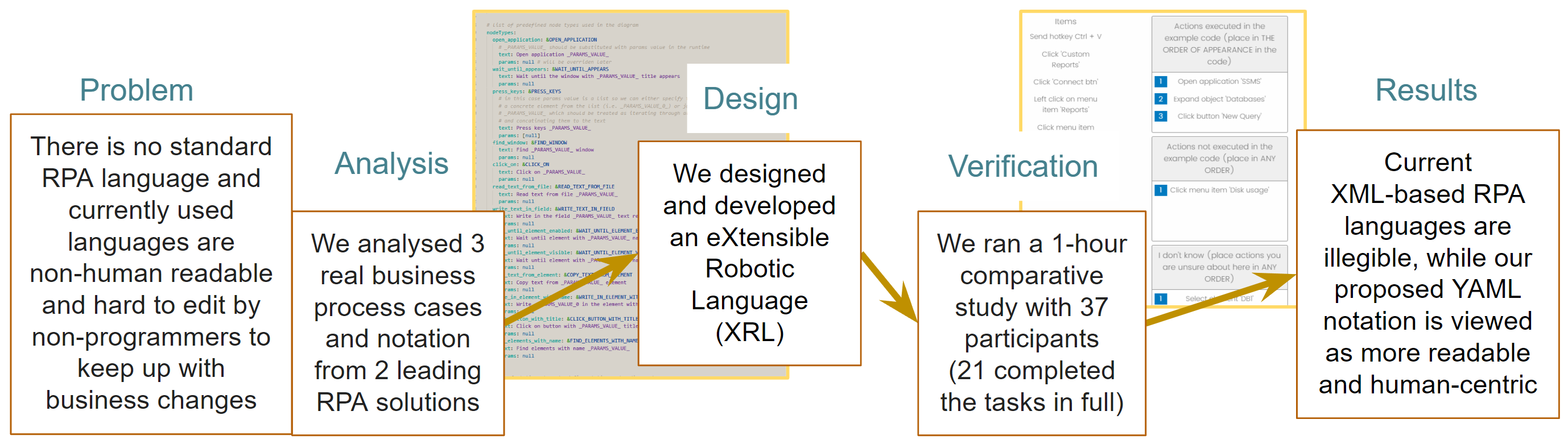}
  \caption{Overview of the study rationale, outputs (with a code sample and study question in the background) and results. Own elaboration.}
  \label{fig:teaser}
\end{figure*}

\section{Introduction}

Motivated by the increasing importance of automating business processes, especially, with Robotic Process Automation (RPA) tools, we set out to address the issues which users may encounter when using RPA. The key problem is that the notation used by the RPA tools differs tool by tool, since there is no single RPA language. Therefore, if users use one tool to automate their business process, they are either stuck with it, or have to pay again, to reimplement their automation in another one. Moreover, the notation used is not human-readable, as it was meant to be used only with GUI solutions of individual providers. This means that users, usually company employees working with repetitive tasks, cannot easily make small changes to the code on their own without engaging with a paid solution. Making the code more human-friendly could empower them to edit and develop automation tasks on their own, without the need to pay for costly market solutions. Such automation should allow employees to increase their qualifications and perform tasks that require more creativity, active decision-making, or interaction with other people, which also tend to be more motivating \cite{willcocks_paper_nodate}. \textbf{The introduction of affordable and approachable RPA tools may eliminate some of the low-quality jobs, allowing employees to perform other more engaging duties} \cite{kopec2018hybrid}. Therefore, in this paper, we verify our proposed unified, human-readable language that would enable employees to become editors of RPA solutions. This would help combat job loss through automation and help break down RPA company monopolies.

\section{Related Work}

IT technologies evolve rapidly and force enterprises to keep transforming their workflows \cite{vedder_challenge_2016}. Among these business processes, only some can be easily automated. These may include interactions with various applications, data manipulation, automatic responses to specific system events or communication with other systems \cite{boulton_what_nodate} such as spreadsheets, Customer Relationship Management (CRM) systems or Enterprise Resource Planning (ERP) software \cite{willcocks_paper_nodate}. Most often, the processes that are best suited for automation have a high volume of transactions, a high level of standardization, well-defined logic, and are mature within the organization structure. Automation of this type of processes brings the most significant financial gains and savings in employee working time (FTE) \cite{hacioglu_handbook_2020}. RPA is also suitable when there is a need to access multiple systems to complete a specific task \cite{hofmann_robotic_2020}. Theoretically, direct integration between two applications can be performed. However, usually, the cost of implementation is too high or integration is not possible due to licensing issues. RPA tools are much cheaper to implement and can provide a similar result to integration \cite{noauthor_rpa_2019,noauthor_rpa_2020}. Therefore, automation of work is becoming more and more common. Robotic process automation (RPA) entered the mainstream around 2015 \cite{welsh_what_nodate} and now, according to Deloitte Global Robotic Survey, 53\% of the companies already use RPA tools \cite{noauthor_global_nodate}. The global market for RPA tools in 2019 was estimated at \$1.4 billion. It is projected to increase to \$11 billion by 2027 \cite{noauthor_robotic_nodate}. Other studies put the real economic value of the potential savings that companies will experience in 2025 at \$5 - \$7 trillion \cite{utermohlen_all_2018}\cite{inc_evolution_nodate}. It is expected that within the next five years, all enterprises will adopt RPA solutions \cite{hofmann_robotic_2020}.

\section{Methods}

\subsection{Study Protocol}

To verify our custom unified human-readable RPA language we built a survey in Qualtrics, with a set of questions providing nonidentifying information about the participants, their language, and their programming background. The key part had three blocks of three questions about three different code samples, generated/written by three different RPA solutions, including our own, representing three different business processes. These were shown to all participants in random order. The tasks were designed to be short so that more people of different backgrounds could solve them.

We informed the participants that they will be asked about their programming experience, but that none was required. We recommend using a large screen device to complete the survey.

\subsubsection{Background Survey} \label{socdat}
Our survey included standard background questions related to age, sex, education level, employment status, and native language of participants. Additionally, we were interested in the declarative level of knowledge of the written English language. We used the Common European Framework of Reference for Languages (CEFR) to generate descriptions of written language proficiency on a 6-point scale (A1-C2).
To learn about the programming experience of the respondents, in addition to a declarative matrix table question on programming skills on a 4-point scale (no experience, beginner, average, advanced) in relation to the five common programming skills (HTML\&CSS, JavaScript, Python, JAVA, C\#, C++, and any other). We added 5 practical questions evaluating the programming background of the participants. The questions are inspired by a study \cite{2021doyoureallycodepreprint} that indicated the top 5 topics to ask to determine if the users have programming experience.

\subsubsection{Comparing the Different Code Samples}

\begin{figure}[h!]
  \centering\includegraphics[width=\linewidth]{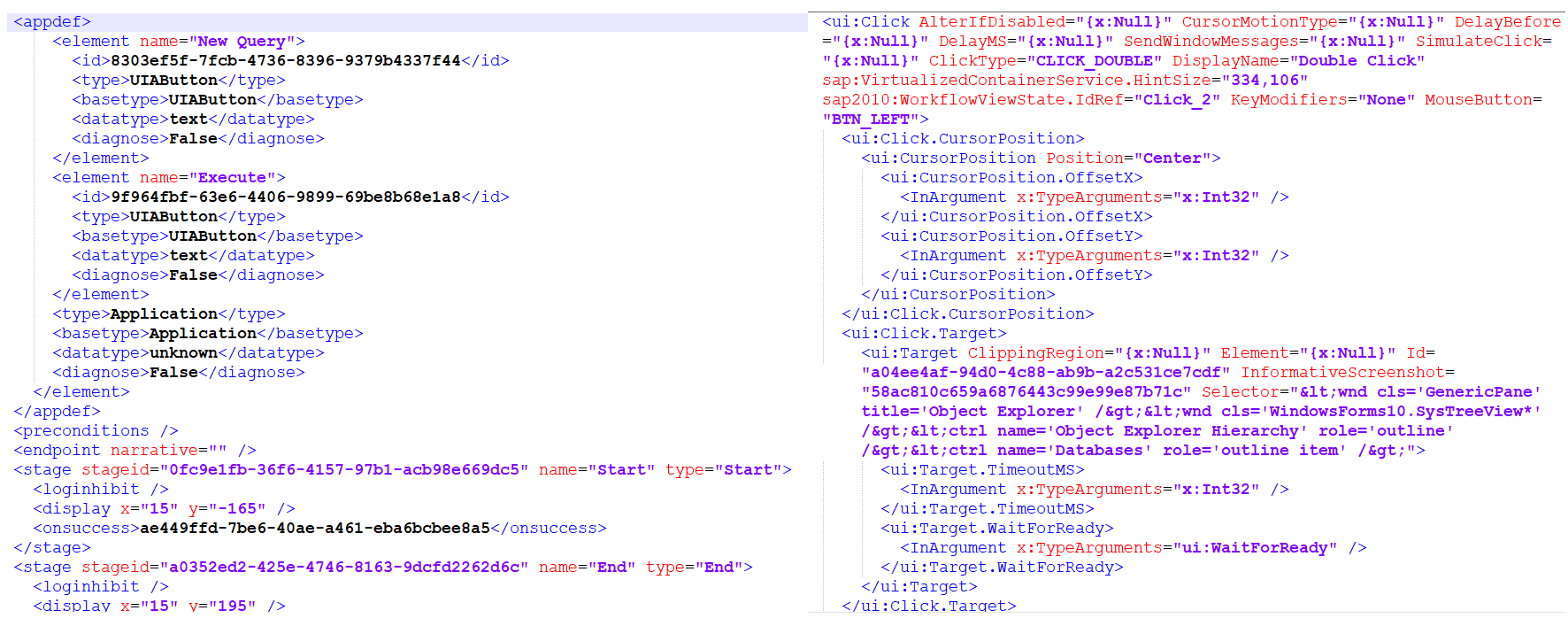}
  \caption{Left: Excerpt from the Code Green sample, showcasing the definition of elements and then actions on them, Right: Excerpt from the Code Blue sample}
  \label{codeblue}
\end{figure}

We provided three code samples based on real business processes. The business processes reflected in the code samples were standardized and trimmed so that they contained the same number of recognizable steps. Each business process had to be sufficiently different so that learning the process would not impact the understanding of the following code samples. Participants could download these samples as a .txt document. These were coded with the names of colors: Blue (157 lines), Green (229 lines), and Red (76 lines).

To unify the examples, we first generated the code using two of the leading RPA solutions. % - BluePrism and UiPath.
We used only basic activities that are familiar to most users, even those without programming experience, such as opening the context menu, pressing shortcuts on the keyboard, or double/single clicks. Later, we removed the parts that were meaningless to the user and lacking a specific context, such as unique identifiers, variables, and attributes needed to find certain elements. Our goal was to make them all the same size in terms of the number of activities. It is worth mentioning that the code example in our language was the only one that was not auto-generated. Each participant was presented with the same samples but in random order. The questions related to the understanding of each sample were all gathered on the same page. First, they had to download the text file that contained the sample code. Next, they were asked to look at the file and at a list of actions and to arrange the actions from the list in the order in which they were executed in the code, decide which ones did not appear in the code, or indicate that they don't know. Finally, after each code sample, participants were asked to express what they thought of each code sample on the Likert scale (Strongly disagree - Disagree - Neither agree nor disagree - Agree - Strongly agree) in the dimensions of:

\begin{itemize}
\item easy, this language in general seems easy to understand
\item human-centric, this language is friendly and descriptive
\item brief, this language is short, elegant with little redundancy
\item systematic, this language is consistent and logical.
\end{itemize}

The dimensions of the above evaluations were based on a review of the literature on the readability and evaluation of programming languages \cite{codereadability,codereadability2020,codereadabilitytesting2016,criteriaforevaluation2016}.  Finally, we asked the respondents to describe their impressions of each of the code samples.

\begin{figure}[h]
  \centering
  \includegraphics[width=0.70\linewidth]{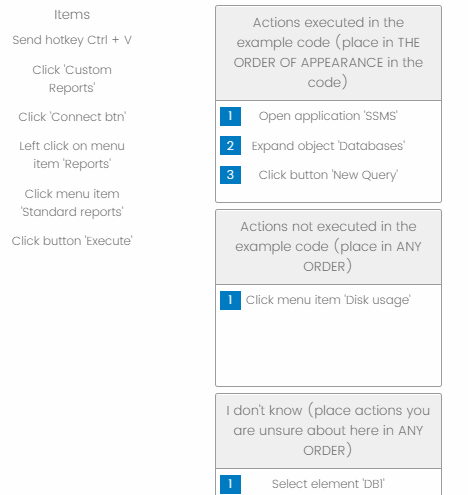}
  \caption{Screenshot of the drag and drop question checking understanding of what happens in the code samples. The questions allowed us to simulate work-life exposure where to modify some functionality, some understanding of the code may be required.}
  \label{draganddrop}
  \vspace{-6mm}
\end{figure}

\section{Results and Discussion}

\subsection{Participants}

This pilot study was conducted between the 1st and 8th of August 2021. The set of participants was recruited from among the authors' coworkers' contacts. Each of the coworkers was asked to send a link to the survey to their friends or family members. In this way, the snowball method was used. Of the 37 people who started the survey (answered at least one question), 21 completed it in full. In the results section, we present only the analyzes performed using the data from the 21 participants, unless otherwise indicated. The average age of the participants was 26, while the median was 23. There were two women and 18 men, while a participant identified as non-binary. Three participants chose C2 as their level of English language, four chose B2, while all others indicated C1 - so the level of English of the participants was generally very good \footnote{For full descriptions of language proficiency (A1-C2) levels see the Council of Europe page: https://www.coe.int/en/web/common-european-framework-reference-languages/level-descriptions}. When it comes to nationality, 2 participants were Ukrainian, one was Russian, and one indicated Russian and Ukrainian, while all of the others were Polish. Most had higher education, either a Master's Degree (8) or Bachelor's Degree (3) while 9 people indicated their highest level of completed education as High School, but out of these 8 were still in school and one was working. Regarding employment, most of the participants had professional experience, 10 were employed, while 6 were studying and employed, only one was seeking opportunities and 4 indicated that they were studying. Most had either good or average programming knowledge, and only 3 had no or very little programming experience. After excluding two outliers, on average, the survey took 45 minutes to complete.

\subsection{Legend and Metrics}

 The described examples will be marked with the names of colors: Blue, Green - solutions from a leading company based on XML, 
%   The process describes opening SQL Server Management Studio and generating a report. 
and Red - our proposed language. For each of the examples, we decided to calculate the percentages related to the degree of completion of the proposed task. Therefore, among the enumerated metrics, we include:
\begin{itemize}
% \scriptsize
  \item \textbf{CO} (correct order) - percentage of correctly identified elements, with their placement in the correct order
  \item \textbf{CU} (correct, unordered) - percentage of correctly identified elements disregarding the order
  \item \textbf{CNU} (correct, not used) - the percentage of correctly identified steps that did not appear in the example. In other words, the person correctly identified the fact that a certain element is not used in the proposed sample.
  \item \textbf{IDK} (don't know) - the percentage of items that were not classified into any of the groups. In other words, the item caused the person a problem and was unable to classify it into any of the before mentioned categories
  \item \textbf{Total completion time in minutes} - time required to complete questions related to the specific code sample
\end{itemize}

% \subsection{Summaries}

% Tables with the aggregated results of the surveys carried out are presented below:

\begin{table}[h!]
\centering
\begin{tabular}{|c|c|c|c|c|c|c|}
\hline
                       ~ & ~ & CO (\%) & CU (\%) & CNU (\%) & IDK (\%) & Total time (min) \\ \hline
\multirow{2}{*}{Blue}  & Average & 28.79 & 59.85 & 60.91 & 21.9 & 15.07 \\
                       & Median  & 25 & 66.67 & 80 & 9.09 & 8.22 \\ \noalign{\hrule height 2pt}
\multirow{2}{*}{Green} & Average & 9.09 & 52.6 & 78.41 & 26.86 & 10.27 \\
                       & Median  & 0 & 57.14 & 100 & 9.09 & 8.8 \\ \noalign{\hrule height 2pt}
\multirow{2}{*}{Red}   & Average & 78.57 & 90.91 & 88.64 & 7.02 & 5.54 \\
                       & Median  & 100 & 100 & 100 & 0 & 5.12 \\ \noalign{\hrule height 2pt}
\end{tabular}
\vspace{2mm}
\caption{Summary of answers for all solutions}
\label{table:summaryAvgMed}
\vspace{-6mm}
\end{table}

\begin{table}[hbt!]
\centering
\begin{tabular}{|c|c|c|c|c|}
\hline
                       ~ & Characteristic & Negative (\%) & Neutral (\%) & Positive (\%) \\ \hline
\multirow{2}{*}{Blue}  & Easy & 69,57 & 8,7 & 21,74 \\
                       & Human-centric  & 73,91 & 8,7 & 17,39 \\
                       & Brief  & 82,61 & 8,7 & 8,7 \\
                       & Systematic  & 43,48 & 34,78 & 21,74 \\ \noalign{\hrule height 2pt}
\multirow{2}{*}{Green}  & Easy & 60,87 & 21,74 & 17,39 \\
                       & Human-centric  & 73,91 & 13,04 & 13,04 \\
                       & Brief  & 69,57 & 26,09 & 4,35 \\
                       & Systematic  & 39,13 & 30,43 & 30,43 \\ \noalign{\hrule height 2pt}
\multirow{2}{*}{Red}  & Easy & 26,09 & 17,39 & 56,52 \\
                       & Human-centric  & 30,43 & 17,39 & 52,17 \\
                       & Brief  & 34,78 & 26,09 & 39,13 \\
                       & Systematic  & 13,04 & 8,7 & 78,26 \\ \noalign{\hrule height 2pt}
\end{tabular}
\vspace{2mm}
\caption{Summary of answers for 5-point Likert scale questions}
\label{table:likertSummaries}
\vspace{-6mm}
\end{table}

\subsection{Example Blue}

Not many people have coped well with placing the items in the correct order. The best result achieved was 83.33\% by one respondent. It was better when we checked only the identification of elements without considering the order. In this case, four respondents achieved a result of 100\%, and nine achieved a result above 67\%. Many people were also reasonably good at identifying elements not in the example. Twelve achieved a result above 73\%.
% Almost every respondent had a problem with identifying at least one element, as evidenced by the results of the IDK metric. 
We also have two results showing that two respondents put all the items in the "Do not know" section. This result correlates with people who indicated their low knowledge of any IT technologies in the previous stages of the survey.
% Most likely, this could lead them to the conclusion that the task is too complex for them.
In the case of the test-taking time, almost all respondents could complete the questions from the Blue example within 30 minutes.

As can be seen from Table \ref{table:summaryAvgMed}, the average correctness achieved by the respondents is about 25\%. However, this score improves if we do not consider the order. In this case, we achieve a result closer to 60\%. 

Summarizing the answers (Table \ref{table:likertSummaries}), we can say that more than 70\% would not describe this language as easy. More than 80\% would not describe this language as human-oriented. More than 90\% would not describe the syntax of the language as concise. When defining the structure as logical, approximately 50\% of the respondents do not agree with this statement. 
% A large part was also unable to say whether they agreed or not.%
Finally, the respondents were asked for a few words about the difficulty of the task and the language itself. Among all the answers, 14 can be classified as definitely unfavorable. The respondents used statements such as "hard to read", "never again", "horrible to read", and "do not have the faintest idea what was that." Several of them mentioned that they could see some logical structure, but they still cannot work out the exact meaning of the example.
% The results obtained show that this example of a language is not best suited for any general use other than automatic processing by a computer.
The syntax does not allow for convenient manual editing. Indeed, the XML format itself adds unnecessary noise and structure that puts a certain cognitive load on the respondents.

\subsection{Example Green}

In this case, the results look worse than in the previous example. Only one respondent was able to achieve a result above 70\% when it comes to metric CO. More than 60\% of the answers related to the CO metric have reached the value of 0\%. Approx. 30\% of the respondents achieved a result of approx—15\%.
% Overall, the average CO metric result is around 9.09\%, and in the case of the median, it is 0\%. 
The CU metric looks a bit better. The mean result and the median here are similar in the range of 50-60\% correct. It seems that it was much easier for the respondents to identify if a given element existed in the proposed example. 
% According to verbal feedback from respondents, this task is often solved with the help of automatic keyword search present in most text editors nowadays. 
30\% of the respondents achieved a result of 0-40\%. Another 50\% achieved a result in the range of 40-78\%. Approx. 20\% of them were able to achieve a result above 78\%. 
% The CNU metric looks best in this example. 
% For most of the respondents, it seems that it is relatively simple to determine which elements are not present in the presented example.
% In this case, over 70\% of the respondents achieved a result above 78\%, many of them reached 100\%.
It appears that, as in the Blue example, it is difficult for most to determine the relationship between the elements used in the examples presented. 

According to Table \ref{table:summaryAvgMed}, the average completion time of Green tasks is 10.27 minutes or 8.80 for the median. In this case, we did not notice large deviations. There is no clear correlation between the duration of the test and the result obtained.
% The correlation between time and the CU metric (we omit CO due to the large number of 0\% results) is 0.2435, which does not indicate a clear correlation. It seems that time is not a big factor in achieving a better result
Then we move on to the results of the 5-point Likert scale relating to the features of the presented language. As can be read from Table \ref{table:likertSummaries}, again, the answers to most questions are "Strongly disagree" and "Disagree." Only the question related to the systematicity of the language seems to polarize the respondents a bit more. Here again, we believe that the same XML format known to most of the respondents has led to the interpretation of this structure as logical. However, the poor results of the CO metrics show that the relationships between the individual elements are not clear to most of the respondents.

The last question was a commentary on the tasks and the example itself. Most of the answers were rather negative, using statements such as "worst language I have ever seen," "I have no idea what it was," and "high redundancy, easy to make mistakes." One respondent pointed out that XML is a somewhat bloated format, making this task much more difficult, especially regarding sequencing items. There were no comments that we could describe as positive.

\subsection{Example Red}
This project design was based on the YAML format. In this case, reducing the XML overhead and making the structure as simple as possible resulted in a much better understanding of the relationships between elements than in the previous examples.% \begin{table}[hbt!]

A substantial proportion of the respondents achieved a result of 100\% for the CO metric. The mean score for this metric was 78.57\%, and the median was 100\%. The highest score was achieved by more than 70\% of the respondents.
% \begin{table}[hbt!]
% \footnotesize
% \centering
% \begin{tabular}{ |c|c|c|c|c|c| } 
% \hline
% & CO (\%) & CU (\%) & CNU (\%) & IDK (\%) &  Time (min) \\
% \hline
% Average & 78.57 & 90.91 & 88.64 & 7.02 & 5.54 \\ 
% Median & 100.00 & 100.00 & 100.00 & 0.00 & 5.12 \\
% \hline
% \end{tabular}
% \vspace{2mm}
% \caption{Summary of answers for the solution Red}
% \vspace{-6mm}
% \end{table}
The value of the CU metric has achieved an even more significant result.% The mean value was 90.90\%, and the median was 100\%. 
In this case, more than 80\% of the respondents achieved the result of 100\%. % In this case, respondents who previously placed all elements in the IDK section also did better.
They had no problem identifying the elements and understanding the sequence of instructions that occurred in succession. 
High values were also achieved for the CNU metric.
% The mean value was 86.63\%, and the median was 100\%. 
After two respondents achieved 25\%, the rest of the results were above 70\%. %The last metric scored lower than in the previous examples. %Almost every respondent put one item in the IDK section. The mean value was 13.63\%, and the median was 9.09\% (one item from the list). %
Here, respondents appeared to be more confident in identifying the individual elements and their purpose within the described process. 
% Task completion times range from a few minutes to a maximum of 10.5 minutes. There are no values ​​that represent a significant deviation. Moreover, the average time needed to complete the task is close to the median - 5-5.5 minutes. 
Then we move on to the results of the 5-point Likert scale question. This time, we can see that most of the answers are within the positive spectrum of values.

We can see that about 60\% of the respondents would describe the language as Easy. More than 50\% would describe the language as being human-oriented. Over 80\% noticed some logic and consistency within the syntax. It was undoubtedly one of the main factors contributing to achieving good CO and CU metrics' results. The answers to the brevity of language (brief) could be more decisive. As we can see here, about 40\% of the respondents answered positively, but at the same time, 36\% replied negatively.
% Here we must remember that we would still like to maintain whole language processability by the computer and the possibility of creating an engine in the future that could be used to automate work on the basis of the proposed syntax.

The last element was the respondents' text answers about the language and overall opinion about the task. In this case, the responses were much more positive. Some people noticed that the syntax is easier to understand than the Blue and Green examples: "I think it's easy to understand," "better than blue and green," "it is ok, I could manage this task without special preparation." 
% As we can see, the respondents noted that they could understand the syntax without any special preparation. 
Some people point out that the syntax seems logical and consistent to them - "the Code seems easy to understand. You need to follow only action name, action parameters, and actions order. For me, this approach is very logical and simple". One of the respondents noticed that the code is a bit lengthy, which results from the need to keep all the elements that would allow us to run a given script with the help of an appropriate engine in the future.
% Finally, some people notice some similarities between the language's syntax and other scripting languages like Python.

\subsection{Limitations}

The evaluation compared a language specifically created for manual editing with existing XML languages; however, this was a necessary compromise to evaluate the existing languages against the one we designed. Moreover, to keep the study below an hour, it was necessary to prepare the tasks and code samples in such a way as to enable even those without programming experience to finish them on time. This is the reasoning behind the use of drag and drop questions, as seen in Figure \ref{draganddrop}. To thoroughly verify if the proposed language syntax is human-centric, code writing, not only comprehension tasks, should be used in further research.

\section{Discussion and Conclusions}
The languages currently used to automate business processes are not suitable for manual editing, as they are illegible to users. They can hardly be used as the basis for a universal language for describing automated processes. They are only used for processing in the selected tools, and it is challenging to imagine editing them outside of such programs. Moreover, they cannot be used to easily migrate the process to a different automating tool.

\begin{table}[hbt!]
\footnotesize
\centering
\begin{tabular}{ |c|c|c|c|c|c| } 
\hline
& CO (\%) & CU (\%) & CNU (\%) & IDK (\%) & Time (min) \\
\hline
Blue & 25.00 & 66.67 & 80.00 & 9.09 & 8.22 \\ 
Green & 0.00 & 57.14 & 100.00 & 9.09 & 8.80 \\ 
Red & 100.00 & 100.00 & 100.00 & 0.00 & 5.12 \\ 
\hline
\end{tabular}
\vspace{2mm}
\caption{Medians for all examples and metrics}
\vspace{-6mm}
\end{table}

The statistics and respondent feedback make it possible to assume that the XML format, most often chosen by the creators of this type of language, is too long and contains too many elements despite its theoretical readability. It makes it particularly difficult to discern the dependencies between individual language elements. A large proportion of the respondents could identify, at least partially, the existence of a given element within the described process. However, respondents needed help to easily identify the item's position as part of the more extensive process described in the example.

Our proposed example (Red) is much more concise than the other examples, and most of the respondents were fine with identifying the elements and the order in which they were processed. The YAML format is a good foundation, offering a consensus between structure, parsability, brevity, and readability. Although we included some elements that will ultimately be needed to process our language in its syntax, they did not affect readability.

\begin{table}[hbt!]
\footnotesize
\centering
\begin{tabular}{ |c|c|c|c|c| } 
\hline
& Easy & Human-centric & Brief & Systematic\\
\hline
Blue & 2.18 & 2.09 & 1.68 & 2.63 \\ 
Green & 2.40 & 2.00 & 2.04 & 2.81 \\ 
Red & 3.45 & 3.31 & 3.18 & 4.00 \\ 
\hline
\end{tabular}
\vspace{2mm}
\caption{Averages from evaluation results}
\vspace{-6mm}
\end{table}

Regarding the evaluation, Red had a higher average score in the Easy and Human-centric categories. A high score was also achieved in all categories. It may show that the adopted structure in the document is legible and does not require additional description. Nevertheless, the need to preserve the computer processability of language requires us to add some additional elements to the language itself, which a human does not necessarily need to interpret a process. For this reason, the syntax was lengthy for some respondents. 
  \vspace{-2mm}
\begin{figure}[h!]
  \centering
  \includegraphics[width=\linewidth]{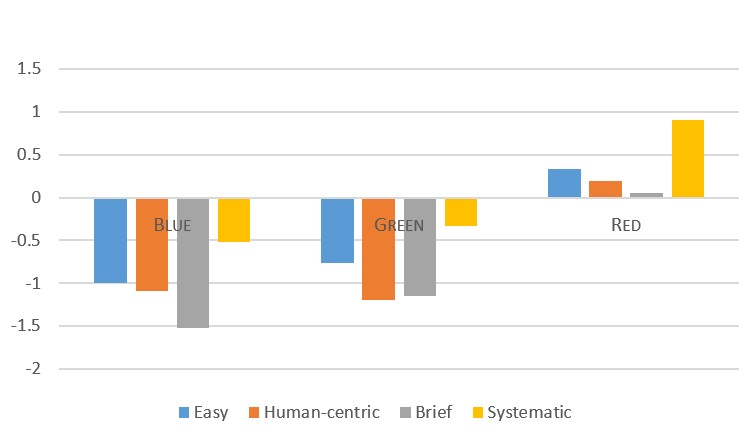}
  \caption{Graphical summary of evaluation results.}
  \vspace{-6mm}
\end{figure}

However, it is worth noting that it is still much shorter than the previous two examples. The descriptive responses of the respondents indicate that Red is characterized by greater simplicity and readability than the other samples. While for nonprogrammers, all the examples were difficult to parse, this was seen as "relatively easy to understand," while one of the respondents, comparing it with other examples, wrote, "never again, but less." This is confirmed by the metrics collected and the overall task execution time, which is also lower than the Blue and Green examples, showcasing its simplicity.

In light of the increased focus on the human aspects of software engineering and well-being in the age of Industry 4.0, research postulating user-centered programming language design is becoming highly relevant. The pervasiveness of technology in every aspect of life and its presence in our professional and private lives make it imperative to empower people to become active creators of the technology solutions that surround them, in this case, software robots, which are expected to become adopted by every industry within the next few years.

\bibliographystyle{splncs04}
\bibliography{bibliography}

\begin{thebibliography}{10}
\providecommand{\url}[1]{\texttt{#1}}
\providecommand{\urlprefix}{URL }
\providecommand{\doi}[1]{https://doi.org/#1}

\bibitem{boulton_what_nodate}
Boulton, C.: What is {RPA}? {A} revolution in business process automation {\textbar} {CIO} (Sep 2018), \url{https://www.cio.com/article/3236451/what-is-rpa-robotic-process-automation-explained.html}

\bibitem{2021doyoureallycodepreprint}
Danilova, A., Naiakshina, A., Horstmann, S., Smith, M.: Do you really code? designing and evaluating screening questions for online surveys with programmers  (03 2021)

\bibitem{noauthor_global_nodate}
Gartner, D.: Global {RPA} {Survey} 2018 {\textbar} {Deloitte} (2018), \url{https://www2.deloitte.com/bg/en/pages/technology/articles/deloitte-global-rpa-survey-2018.html}

\bibitem{noauthor_rpa_2019}
Gartner, D.: {RPA} vs integration: {The} differences between the solutions (Sep 2019), \url{https://digitalworkforce.com/rpa-news/robotic-process-automation-vs-integration-2/}

\bibitem{noauthor_rpa_2020}
Gartner, D.: {RPA} vs. {Systems} {Integration} (Sep 2020), \url{https://smartbridge.com/rpa-vs-systems-integration/}

\bibitem{noauthor_robotic_nodate}
Gartner, D.: Robotic {Process} {Automation} {Market} {Size} {Report}, 2021-2028 (Apr 2021), \url{https://www.grandviewresearch.com/industry-analysis/robotic-process-automation-rpa-market}

\bibitem{hacioglu_handbook_2020}
Hacioglu, U., Lee, I. (eds.): Handbook of {Research} on {Strategic} {Fit} and {Design} in {Business} {Ecosystems}:. Advances in {E}-{Business} {Research}, IGI Global (2020). \doi{10.4018/978-1-7998-1125-1}, \url{http://services.igi-global.com/resolvedoi/resolve.aspx?doi=10.4018/978-1-7998-1125-1}

\bibitem{hofmann_robotic_2020}
Hofmann, P., Samp, C., Urbach, N.: Robotic process automation. Electronic Markets  \textbf{30}(1) (Mar 2020). \doi{10.1007/s12525-019-00365-8}, \url{http://link.springer.com/10.1007/s12525-019-00365-8}

\bibitem{inc_evolution_nodate}
Inc, U.: The {Evolution} of {Robotic} {Process} {Automation} ({RPA}): {Past}, {Present}, and {Future} (Jul 2016), \url{https://www.uipath.com/blog/rpa/the-evolution-of-rpa-past-present-and-future}

\bibitem{kopec2018hybrid}
Kope{\'c}, W., Skibi{\'n}ski, M., Biele, C., Skorupska, K., Tkaczyk, D., Jaskulska, A., Abramczuk, K., Gago, P., Marasek, K.: Hybrid approach to automation, rpa and machine learning: a method for the human-centered design of software robots. arXiv preprint arXiv:1811.02213  (2018)

\bibitem{codereadability}
Oliveira, D., Bruno, R., Madeiral, F., Castor, F.: Evaluating code readability and legibility: An examination of human-centric studies. In: 2020 IEEE International Conference on Software Maintenance and Evolution (ICSME). pp. 348--359 (2020). \doi{10.1109/ICSME46990.2020.00041}

\bibitem{codereadabilitytesting2016}
Sedano, T.: Code readability testing, an empirical study (04 2016). \doi{10.1109/CSEET.2016.36}

\bibitem{criteriaforevaluation2016}
Sheikh, G., Islam, N.: A qualitative study of major programming languages: teaching programming languages to computer science students. International Journal of Information and Communication Technology  (01 2016)

\bibitem{codereadability2020}
Tariq, M.U., Bashir, M., Babar, M., Sohail, A.: Code readability management of high-level programming languages: A comparative study. International Journal of Advanced Computer Science and Applications  \textbf{11},  595--602 (03 2020). \doi{10.14569/IJACSA.2020.0110375}

\bibitem{utermohlen_all_2018}
Utermohlen, K.: All the {Robotic} {Process} {Automation} ({RPA}) {Stats} {You} {Need} to {Know} (Apr 2018), \url{https://towardsdatascience.com/all-the-robotic-process-automation-rpa-stats-you-need-to-know-bcec22eaaad9}

\bibitem{vedder_challenge_2016}
Vedder, R., Guynes, C.S.: The {Challenge} {Of} {Botsourcing}. Review of Business Information Systems (RBIS)  \textbf{20}(1) (May 2016). \doi{10.19030/rbis.v20i1.9677}, \url{https://clutejournals.com/index.php/RBIS/article/view/9677}

\bibitem{welsh_what_nodate}
Welsh, J.: What the {History} of {RPA} {Technology} {Says} {About} its {Future} (Mar 2019), \url{https://globalpayrollassociation.com/blogs/technology/what-the-history-of-rpa-technology-says-about-its-future}

\bibitem{willcocks_paper_nodate}
Willcocks, P.L.: Paper 15/05 {The} {IT} {Function} and {Robotic} {Process} {Automation} p.~39 (Oct 2015)

\end{thebibliography}

\end{document}